\documentstyle[11pt,aaspp4]{article}

\newcommand{\vf}{\rm}
\newcommand{\et}{{et al}\ }
\newcommand{\F}{{\rm F}}
\newcommand{\half}{{\textstyle{\frac{1}{2}}}}

\newcommand{\I}{{\scriptscriptstyle I}}

\newcommand{\V}{{\scriptscriptstyle V}}
\newcommand{\X}{{\scriptscriptstyle X}}
\newcommand{\Y}{{\scriptscriptstyle Y}}
\newcommand{\RM}{{\rm RM}}
\newcommand{\LDM}{{\rm LDM}}
\newcommand{\phiav}{\langle\phi_\F\rangle}
\begin{document}

\title{Stochastic Faraday Rotation}

\author{D.B. Melrose$^{1}$ and J.-P. Macquart$^{2}$\\
Research Centre for Theoretical Astrophysics\\
School of Physics, University of Sydney\\
$^{1}$melrose@physics.usyd.edu.au,
$^{2}$jpm@physics.usyd.edu.au}

\date{\today}


\begin{abstract}

Different ray paths through a turbulent plasma can produce
stochastic Faraday rotation leading to depolarization of any
linearly polarized component. Simple theory predicts that the
average values of the Stokes parameters decay according to
$\langle Q\rangle$, $\langle U\rangle\propto\exp(-\delta_l)$,
with $\delta_l\propto\lambda^4$. It is pointed out that a definitive test
for such depolarization is provided by the
fact that $\langle Q^2+U^2\rangle$ remains constant while
$\langle Q\rangle^2+\langle U\rangle^2$ decreases
$\propto\exp(-2\delta_l)$. The averages to which this effect,
called polarization covariance, should apply are discussed;
it should apply to spatial averages over a polarization map
or temporal averages over a data set, but not to beamwidth
and bandwidth averages that are intrinsic to the observation
process. Observations of depolarization would provide
statistical information on fluctuations in the turbulent
plasma along the line of sight, specifically, the variance of
the rotation measure. Other effects that can also cause
depolarization are discussed. Favorable data sets to which
the test could be applied are spatial averages over
polarization maps of the lobes of radiogalaxies and temporal
averages for the binary pulsar PSR B1259-63 as it goes into
and comes out of eclipse by the wind of its companion.

\noindent
{\bf Key words: Faraday rotation, turbulence,
magnetic fields}

\end{abstract}


\section{Introduction}

Faraday rotation can cause depolarization of an astrophysical
synchrotron source in three ways: internal depolarization,
bandwidth depolarization and external depolarization
(e.g., Burn 1966). Internal
depolarization is due to differential Faraday rotation
when a ray emitted at the far side of the source has its plane
of polarization rotated through $\ga\pi/2$ before reaching
the near side of the source. Bandwidth depolarization occurs
due to differential rotation of the plane of polarization
across the bandwidth of observation (e.g., Simonetti,  Cordes
\& Spangler 1984; Lazio, Spangler \&
Cordes 1990). External depolarization is
due to a random component of the magnetic field along the ray
path between the source and the observer through the
intervening medium (interplanetary, interstellar,
intergalactic or intracluster).  Two different treatments of
external depolarization are available. One treatment,
referred to here as stochastic Faraday rotation, is based on
the assumption that the (Faraday) angle through which the
plane of polarization is rotated contains a random component.
(e.g., Burn 1966; Simonetti \et 1984; Tribble 1991). The
other treatment of external depolarization involves including
the effect of the magnetic field in scintillation theory
(Erukhimov \& Kirsh 1973; Tamoikin \&
Zamek 1974; Melrose 1993a,b [hereinafter
papers~A and~B]). Simonetti \et (1984) established that these
two treatments are equivalent for the mean values of the
Stokes parameters, as shown in the Appendix below. Such
stochastic Faraday rotation is of practical interest in
determining the properties of the medium through which the
radiation propagates: measurement of the frequency at which
depolarization occurs provides information on the
fluctuations in the magnetic field along the line of sight
which complements information from the dispersion measure
(DM), rotation measure (RM) and scattering measure (SM),
e.g., Taylor \& Cordes (1993). However, in cases where
depolarization is observed it is not necessarily clear that
it is due to stochastic Faraday rotation, rather than to one
of the other depolarizing mechanisms.

In the present paper, it is shown that a definitive test for
stochastic Faraday rotation follows by measuring the mean
square values of the Stokes parameters, specifically
$\langle Q^2\rangle$, $\langle U^2\rangle$,
$\langle QU\rangle$. These mean values can be derived in a
very simple way by assuming that the Faraday rotation angle
can be treated as a random variable. Let the Faraday angle be
\begin{equation}
\label{eq1.1}
\phi_\F=\phi_0+\RM\,\lambda^2,
\end{equation}
where $\phi_0$ is the angle at the source and $\lambda$ is the
wavelength. The statistical average is performed by assuming
that $\phi_\F$ has a random component that obeys gaussian
statistics. This simple model is justified in an Appendix,
where it is shown that the theory of scintillations (e.g. Narayan 1992) in
a magnetized plasma leads to the same result for
$\langle Q^2\rangle$, $\langle U^2\rangle$,
$\langle QU\rangle$ as this model. The model implies that
$\langle Q^2+U^2\rangle$ is a constant, that is, although
$\langle Q\rangle$, $\langle U\rangle$ decay,
$\langle Q^2+U^2\rangle$ does not. This simple but
surprising result is referred to here as polarization
covariance; it distinguishes depolarization due to stochastic
Faraday rotation from other depolarization mechanisms.

In section~2 a model for stochastic Faraday rotation is
used to derive formulae that describe the depolarization and
the polarization covariance. In section~3, the nature of the
averaging process is discussed. In section~4, some specific
examples of possible depolarization are considered.
The conclusions are summarized in section~5.

\section{Random Variations in the Rotation Measure}

Faraday rotation results from the difference between the
refractive indices of the two natural modes of a magnetized
plasma. In the simplest approximation the polarization of the
natural modes is assumed circular, and then Faraday rotation
affects only the linear polarization. Here the polarization
is described in terms of the Stokes parameters, $I$, $Q$,
$U$, $V$. Only $Q$, $U$ are affected by Faraday rotation; $I$,
$V$ are not mentioned further in this paper.

\subsection{Faraday rotation}
Faraday rotation causes the Stokes parameters $Q$, $U$ to
vary with distance $z$ along the ray path according to
{
\catcode`#\active
\setbox0=\hbox{$-$}\dimen0=\wd0
\def #{\hbox to \dimen0{\hfil}}
\begin{equation}
\label{eq2.1}
\frac{\partial}{\partial z}\,
\pmatrix{{Q}(z)\cr{U}(z)\cr}
=\rho_\V(z)\pmatrix{
0&-1\cr
1&#0\cr}
\pmatrix{{Q}(z)\cr{U}(z)\cr},
\label{eq2.2}
\end{equation}
\begin{equation}
\rho_\V
=\frac{e^3}{8\pi^2\varepsilon_0m_e^2c^3}\,
n_eB_\parallel\,\lambda^2.
\end{equation}

}

\noindent It is convenient to write equation (\ref{eq2.1}) in the form
(e.g., Spangler 1982)
\begin{equation}
\label{eq2.3}
\frac{\partial}{\partial z}\,{\cal P}(z)
=i\rho_\V(z)\,{\cal P}(z),
\quad
{\cal P}=Q+iU.
\end{equation}
The solution of equation (\ref{eq2.3}) is
\begin{eqnarray}
\label{eq2.4}
{\cal P}(z)&=&e^{i\phi_\F(z)}{\cal P}_0,
\nonumber\\
\phi_\F(z)&=&\int_0^zdz'\,\rho_\V(z'),
\end{eqnarray}
where ${\cal P}_0=Q_0+iU_0$ is determined by the
intrinsic polarization at the source (at $z=0$). In the following it is assumed that the axes are oriented such that $U_0=0$, so
that one has ${\cal P}_0=Q_0$.

\subsection{Fluctuations in the Faraday angle}
Fluctuations in the plasma parameters along the ray path may
be taken into account by separating the Faraday angle,
$\phi_\F$, into an average value, $\phiav$, and a
fluctuating part, $\delta\phi_\F$:
\begin{equation}
\label{eq2.5}
\phi_\F=\phiav+\delta\phi_\F, \quad
\phiav=\langle\rho_\V\rangle L,
\end{equation}
where $L$ is the length of the ray path.
It is convenient to incorporate the average
Faraday rotation into modified Stokes parameters, denoted by
tildes (paper~A):
\begin{equation}
\label{eq2.6}
{\tilde{\cal P}}(z)=e^{-i\phiav}{\cal P}(z)
=e^{i\delta\phi_\F(z)}Q_0,
\end{equation}
where the final equality follows from equation (\ref{eq2.4}).
One has ${\tilde{\cal P}}={\tilde Q}+i{\tilde U}$, with
${\tilde Q}$, ${\tilde U}$ real. The
fluctuating part of the Faraday angle,
$\delta\phi_\F$, on the right hand side of equation (\ref{eq2.6}) is
now assumed to be a random variable.

\subsection{The mean Stokes parameters}
The mean values of the Stokes parameters follow from the
statistical average of equation (\ref{eq2.6}), which may be evaluated
using the property,
$\langle\exp(i\phi)\rangle=\exp(-\langle\phi^2\rangle/2)$ for
a gaussian random variable, $\phi$. (The derivation of this
property follows from the power series expansion of the
exponential and $\langle\phi^{(2n+1)}\rangle=0$,
$\langle\phi^{2n}\rangle=3\cdot5\cdot\ldots\cdot(2n-1)
\langle\phi^2\rangle^n$ for $n\ge0$ and integer.) This gives
\begin{equation}
\label{eq2.7}
\langle{\tilde{\cal P}}\rangle=Q_0
\exp[-\half\langle(\delta\phi_\F)^2\rangle].
\end{equation}
It follows that the Stokes parameters vary
according to
\begin{eqnarray}
\label{eq2.8}
\langle{\tilde Q}\rangle&=&Q_0e^{-\delta_l},
\quad
\langle{\tilde U}\rangle=0,
\end{eqnarray}
where equation (\ref{eq1.1}) is used and $\delta_l = \langle (\delta
\phi_F)^2\rangle/2$. The decay of $\langle{\tilde
Q}\rangle$ described by equation (\ref{eq2.8}) corresponds to a
depolarization of the radiation. This result has been derived
both by the foregoing method (Burn 1966, Tribble
1991) and also from  scintillation theory for
a magnetized plasma in the special case of axial rays
(Erukhimov \& Kirsh 1973; Tamoikin \&
Zamek 1974; Simonetti \et 1984; paper~A). The exponent of the
decay factor in equation (\ref{eq2.8}) is determined by
\begin{equation}
\langle(\delta\phi_\F)^2\rangle=
\int_0^Ldz\int_0^Ldz'\,
\langle\delta\rho_\V(z)\,\delta\rho_\V(z')\rangle,
\hfill
\label{dphi2}
\end{equation}
with
$\delta\rho_\V=\rho_\V-
\langle\rho_\V\rangle$. Equation (\ref{dphi2}) may be
rewritten by analogy with equation (\ref{eq2.5}) in the form
$\langle(\delta\phi_\F)^2\rangle
=\langle\delta\rho_\V^2\rangle L^2$.

\subsection{The mean square Stokes parameters}
The mean values of the squares of the Stokes parameters and of the
products of the Stokes parameters follow from the statistical average of
the modulus squared of (\ref{eq2.6}) and of the product of equation
(\ref{eq2.6}) and its complex conjugate.  Under the assumption that
$\delta \phi_F$ is a gaussian random variable, these give
\begin{eqnarray}
\label{eq2.10}
\langle{\tilde{\cal P}}^2\rangle&=&Q_0^2
\exp(-4\delta_l),
\nonumber\\
\langle{\tilde{\cal P}}{\tilde{\cal P}}^*\rangle&=&Q_0^2.
\end{eqnarray}
In terms of the Stokes parameters, equation (\ref{eq2.10}) implies
\begin{eqnarray}
\label{eq2.11}
\langle{\tilde Q}^2\rangle&=&Q_0^2
e^{-2\delta_l}\cosh(2\delta_l),
\nonumber\\
\langle{\tilde U}^2\rangle&=&Q_0^2
e^{-2\delta_l}\sinh(2\delta_l),
\nonumber\\
\langle{\tilde Q}{\tilde U}\rangle&=&0.
\end{eqnarray}
The result (\ref{eq2.11}) was derived in
paper~B by solving the fourth order moment equation
for a magnetized plasma for axial rays. The
fact that the foregoing theory reproduces the result of the
more general theory confirms the equivalence of the two
theories for the higher order moments for axial rays.

The covariances of the polarized component of the radiation
are given by
\begin{eqnarray}
\label{eq2.12}
\frac{\langle{\tilde Q}^2\rangle}
{\langle{\tilde Q}\rangle^2}
&=&\cosh(2\delta_l),
\nonumber\\
\frac{\langle{\tilde U}^2\rangle}
{\langle{\tilde Q}\rangle^2}
&=&\sinh(2\delta_l).
\end{eqnarray}
It follows that for small $\delta_l$ the mean
square of ${\tilde Q}$ is equal to the square of its mean;
however, as $\delta_l$ increases, the ratio of the mean square
fluctuations to the square of the mean increases, becoming
arbitrarily large for $\delta_l\gg1$. The mean of
${\tilde U}$ is zero by definition, and the mean square of
${\tilde U}$ is small compared to the mean square of
${\tilde Q}$ for small $\delta_l$, but one has
$\langle{\tilde Q}^2\rangle
\approx\langle{\tilde U}^2\rangle$ for
$\delta_l\gg1$. This somewhat surprising prediction may be
attributed to the second of equations (\ref{eq2.10}): because the
Faraday angle does not appear, the value of $Q^2+U^2$ is
independent of the statistical averaging. Whether or not
this is the case for averaging processes in practice is
discussed in the next section.

\subsection{Consequences of polarization covariance}
The relations (\ref{eq2.7}), (\ref{eq2.10}) and  may be
rewritten in a variety of ways. The  tildes on the Stokes
parameters correspond to, cf.\ equation (\ref{eq2.6})
\begin{eqnarray}
\label{tildes}
{\tilde Q}&=&Q\cos\langle\phi_\F\rangle
+U\sin\langle\phi_\F\rangle,
\nonumber\\
{\tilde U}&=&-Q\sin\langle\phi_\F\rangle
+U\cos\langle\phi_\F\rangle.
\end{eqnarray}
The relations (\ref{eq2.11}) involve the parameters
$\langle\phi_\F\rangle$ and $\delta_l$, but not $Q_0$. Various
relations between the averages of the Stokes parameters
(without the tildes) may be derived by using equation (\ref{tildes})
and eliminating one or more of these three parameters. For
example, relations that involve $Q_0$, $\delta_l$, but not
$\langle\phi_\F\rangle$, are
\begin{equation}
\label{polcov}
\langle Q^2\rangle+\langle U^2\rangle
-\langle Q\rangle^2-\langle U\rangle^2
=Q_0^2(1-e^{-2\delta_l}),
\end{equation}
\begin{equation}
\langle Q^2\rangle\langle U^2\rangle
-\langle QU\rangle^2={\textstyle{1\over4}}
Q_0^4(1-e^{-8\delta_l}).
\end{equation}
More generally, there are five observable parameters in this
theory, $\langle Q\rangle$, $\langle U\rangle$,
$\langle Q^2\rangle$, $\langle U^2\rangle$,
$\langle QU\rangle$, and three unknowns,
$\langle\phi_\F\rangle$, $\delta_l$, $Q_0$. Thus measurement
of all five observables would provide two constraints on the
theory. One of these is equivalent to
$\langle{\tilde U}\rangle=0$ implying
$\langle{\tilde Q}{\tilde U}\rangle=0$, and the other is a
definitive test for stochastic Faraday rotation.

\section{Conditions for Polarization Covariance to be
Observed}

The conditions under which polarization covariance should be
observed may be identified by considering the manner in
which the observational data are recorded and the nature of
the averaging process intrinsic in, or applied to, the data.
For a given radiotelescope observing at a given frequency,
data are recorded over an integration time, $\Delta t$,
and over a number of frequency channels, each of bandwidth,
$\Delta\nu$, centered on the given frequency. In
principle, the basic bits of information are the values of
$I$, $Q$, $U$ (and maybe $V$) for each $\Delta t$ and
$\Delta\nu$. Over an observation time, the averages of $I$,
$Q$, $U$ are simply the averages of the values for each
$\Delta t$, and similarly the values of $I^2$, $Q^2$, $U^2$,
$QU$, etc., are to be obtained by storing the appropriate
products for each $\Delta t$ and then averaging each of them
over the observation time. The data may also be averaged in
frequency over all the channels. Furthermore, the telescope
has a characteristic beamwidth, $\Delta\theta$, and the
bits of data are already averaged over this beamwidth. Hence, one may identify three types of average: intrinsic
averages in the data over $\Delta t$, $\Delta\nu$,
$\Delta\theta$; the averages applied to the data
over the observation time and bandwidth; and temporal
averaging over times long compared with a given observation
or spatial averages on scales large compared with
the resolution of the telescope.

\subsection{Averages over space or time}

The interpretation of an ensemble average, denoted by the
angular brackets above, is not specified a priori. It should
apply to the averages over space, time, or beamwidth. As the
space scale or time scale over which the average is performed
is increased, the value of $\delta_l$ should increase (or
remain constant). In principle, measurement of $\delta_l$
provides information on the fluctuations over the relevant
time or space scales.

Spatial averaging requires a map showing the variation of RM
(e.g., Simonetti \et 1984, Simonetti \& Cordes 1988). To
simplify the discussion, suppose that the Faraday rotation
all occurs in a screen at a distance $L$. An angular
separation, $\delta\theta$, corresponds to a transverse
displacement $r=L\delta\theta$ at the screen, and hence an
average over $\delta\theta$ corresponds to an average over
transverse displacement at the screen. Simonetti \et (1984)
introduced a structure function for RM:
\begin{equation}
\label{eq3.1}
\kappa_\RM(\delta\theta)=
\langle[\RM(\theta)-\RM(\theta+\delta\theta)]^2\rangle,
\end{equation}
and showed that it consists of a statistical part and a
geometric part. The statistical part for a power law spectrum of
fluctuations is of the form
\begin{equation}
\label{eq3.2}
\kappa_\RM(\delta\theta)\propto(\delta\theta)^\beta
\end{equation}
for $0<\beta<2$, with $\beta=5/3$ corresponding to a
Kolmogorov spectrum. The geometric part is due to the
different path lengths through the random medium, and it
appears to be unimportant in practice.

The condition for the theory of stochastic Faraday rotation
to apply is that the variations in the phases of the
components in the two modes be large, so that their
difference, and hence the Faraday angle, may be treated as a
random variable (Lee \& Jokipii 1975; Simonetti \et 1984).
The phase difference between different points would be known
for an ideal map, but in practice the sampling is coarse and
the separation over which phase coherence is lost is not
known a priori. The theory for depolarization, as described by
equations (\ref{eq2.7}) and (\ref{eq2.11}), applies only over
separations large compared with that over which phase
coherence is lost.

Temporal variations in the Faraday angle can arise from two
effects: intrinsic temporal changes in the screen (e.g., due
to wave motions therein), and convective motions, involving
relative motions between the the screen and the observer.
Only convective motions are considered here. An angular
displacement, $\Delta\theta$, of the image then occurs in a
time interval, $\Delta t$, related by
\begin{equation}
\label{eq3.8}
L\Delta\theta=v\Delta t,
\end{equation}
as the pattern sweeps across the observer at speed $v$.

For an unresolved source with an apparent size $\theta_{\rm eff}$,
the minimum time on which different ray paths are sampled is
\begin{equation}
\label{eq3.9}
t_{\rm min}={\rm min}\,[\theta_{\rm eff},\theta_B]L/v,
\end{equation}
and temporal averaging is relevant only over times $\gg t_{\rm min}$.
For a resolved source that fills the beam the minimum time for
temporal averaging is set by the beamwidth of the telescope, $t_{\rm
min}=L\theta_B/v$. By way of illustration, for a screen at a
distance of several kiloparsecs and with $v\sim100\rm\,km\,s^{-1}$
(of order the velocity of the Earth through the Galaxy), even for a
source of apparent angular size
$\theta_{\rm eff}\sim1\rm\,mas$, equation (\ref{eq3.9}) implies that
$t_{\rm min}\sim2$ months is the minimum time for a temporal
average to show the effect of stochastic Faraday rotation for
observations with milliarcsecond resolution. The
time scale for temporal variations is shorter for
observations with higher angular resolution and for closer
screens. The effect is of no practical interest for
extragalactic screens.

\subsection{Beamwidth averaging}

Beamwidth averaging is relevant for an extended, resolved
source, that is, for a source with angular size, $\theta_S$,
satisfying $\theta_S>\theta_B$. A source with
$\theta_S\ll\theta_B$ is unresolved and beamwidth averaging is
not relevant.

The observed intensity for a resolved source is the actual
intensity convolved with a point spread function,
$p({\bf r})$, where ${\bf r}$  is the perpendicular
displacement on an image screen  (e.g., Rohlfs \& Wilson
1996, Eqn.\ (5.64)). For a gaussian beam one has
\begin{equation}
p({\bf r})={1\over2\pi r_0^2}\,
\exp(-r^2/{2r_0}^2),
\label{psf}
\end{equation}
where $r_0$ is a constant.

Consider an idealized model in which
$\delta\phi_\F({\bf r},z)$ varies due to a wave, with
wavenumber $K$ and amplitude $\delta B$, propagating along
the $x$-axis with magnetic fluctuations along the $z$-axis.
Specifically, assume
$\delta\phi_\F({\bf r},z)=f_0\sin Kx$, with
$f_0=(\delta B/B_\parallel)\RM\lambda^2$. If the beamwidth
were zero one would observe
${\cal P}=Q_0\exp(if_0\sin Kx)$. On convolving this with the
point spread function (\ref{psf}), one finds
\begin{eqnarray}
\label{eq3.5}
\langle{\tilde{\cal P}}\rangle_b=
Q_0\sum_{n=-\infty}^\infty J_n(f_0)\,
e^{inKx}\,
e^{-n^2K^2r_0^2/2},
\end{eqnarray}
where the subscript $b$ denotes the average over beamwidth.
In the limit $r_0\to0$ the final exponential in equation (\ref{eq3.5})
goes to unity, and the sum gives
${\cal P}=Q_0\exp(if_0\sin Kx)$, as required. For $Kr_0\gg1$,
$\langle{\tilde{\cal P}}\rangle_b$ decreases sharply and
only the term $n=0$ contributes to the sum in equation (\ref{eq3.5}).

Although this model is highly idealized, it suffices to
illustrate that significant depolarization occurs for
$Kr_0\gg1$, that is, when the fluctuations in RM occur over a
scale, $\sim1/K\ll r_0$, that is small compared with the
resolution of the telescope.

For an unresolved source, $\theta_S<\theta_B$, the
beamwidth average is replaced by an average over the
apparent source. The apparent size cannot be smaller than the
scatter-image size of a point source affected by scattering
in the interstellar medium (ISM), or other turbulent medium
through which the radiation passes (e.g., Rickett 1990). A
simple model for the angular broadening due to such scattering
(e.g., Blandford \& Narayan 1985) involves a gaussian profile
similar to that assumed above in performing the beamwidth
average. Now $I({\bf r})\propto\exp(-{r}^2/{2r_1}^2)$ is
interpreted as the brightness distribution as a function of
${\bf r}$ for an apparent source of angular size $\sim
r_1/L$. Thus, for a point source the average over
beamwidth is effectively that over the angular size of the
scatter image, rather than over the beamwidth of the
radiotelescope.

\subsection{Average over bandwidth}

An average over bandwidth causes depolarization because
$\phi_\F\propto\nu^{-2}$ is a function of frequency. Over
$\Delta\nu$ this causes a change
$\vert \Delta\phi_\F \vert = \vert 2\phi_\F\Delta\nu/\nu \vert$, and hence
a fractional reduction in the degree of linear polarization by
$2\Delta\nu/\nu$. Depolarization due to bandwidth averaging
is qualitatively different from the other averages considered
above because it affects the amplitude rather than the
intensity. Polarization covariance applies to the statistics
of the measurements of $Q$ and $U$, and bandwidth averaging
limits the ability to measure values of $Q$ and $U$.

\section{Linear depolarization measure (LDM)}

Assuming that depolarization is observed due to stochastic
Faraday rotation in the ISM, the wavelength at which the
depolarization becomes important provides information on the
statistical properties of the ISM. This may be described in
terms of a `linear depolarization measure', LDM, defined, cf.\ equation (\ref{eq2.8}),
\begin{eqnarray}
\label{ldm}
\delta_l&=&\half\langle(\delta\phi_\F)^2\rangle
=\half\LDM \lambda^4,
\nonumber\\
\LDM
&=&
\langle(\RM-\langle\RM\rangle)^2\rangle,
\end{eqnarray}
where equation (\ref{eq1.1}) is used. The quantity LDM (linear
depolarization measure) is determined by the variance in RM,
and may be measured by any observation of stochastic
depolarization.

Four observationally determined parameters have been used to
describe the statistical properties of the ISM: the
dispersion measure, DM, the emission measure EM, the
scattering measure SM and the rotation measure RM (e.g.,
Cordes \& Lazio 1991; Cordes \et 1991). These parameters
determine the integral along the line of sight of,
respectively, $n_e$, $n_e^2$, $C_N^2$ and $n_eB_\parallel$,
where $C_N^2$ is the structure constant for the electron
density fluctuations. The additional parameter, LDM, is
determined by the integral along the line of sight of
$(n_eB_\parallel- \langle n_eB_\parallel\rangle)^2$.

To illustrate how the LDM could be used to infer information
concerning fluctuations in the magnetic field, consider two
extreme cases (e.g., Simonetti \& Cordes 1988). On the one
hand, if the magnetic field in the ISM is relatively uniform,
then the fluctuations in $(n_eB_\parallel)^2$ are dominated
by those in $n_e^2$, and one would expect the ratio
LDM/RM$^2$ to vary across the Galaxy in the same way as does
the ratio SM/DM$^2$. On the other hand, if the magnetic field
is sufficiently random, or if regions of opposite sign of
$B_\parallel$ nearly cancel, then one can have $\langle
B_\parallel^2\rangle\gg \langle B_\parallel\rangle^2$,
implying that LDM/\RM$^2$ could exceed unity, due to near
cancellation of contributions to RM of opposite sign along
the line of sight.

Existing data suggest that the variance in RM in the ISM is of
the same order but not much larger than $\langle \RM\rangle^2$
(Simonetti \et 1984; Simonetti \& Cordes 1988). The
implication is that $B_\parallel$ does not change sign many
times along these lines of sight. Further data confirming (or
otherwise) this result is clearly important for our
understanding of the structure of the magnetic field in the
ISM. In principle, polarization covariance can be used to
estimate $\langle \RM\rangle^2$ with much finer resolution
than can be achieved by averaging over a polarization map.

\section{Observations of depolarization}

Depolarization observed in some sources has been interpreted
in terms of stochastic Faraday rotation, but the definitive
test for polarization covariance has not been performed. In
this section, some specific examples are discussed.

\subsection{Depolarization in the ISM}

Lazio \et (1990) used polarization data for eight
double-lobed radio galaxies to discuss variations in RM in
the Cygnus region. They reported significant depolarization,
for several of these sources, at $\lambda\ga10\rm\,cm$. For
sources for which RM could be determined, values in the range
$\sim250$--$800\rm\,rad\,m^{-2}$ were found. The data were
compared with a model that gives a relation of the form, cf.\
equation (\ref{eq3.2}),
$\kappa(\theta)=\kappa_0
(\theta/\theta_0)^{\beta}$,
with $\kappa_0\sim(10\rm\,rad\,m^{-2})^2$ for
$\theta_0\sim10''$ and with $\beta=5/3$. The data are
described moderately well by the model, but a smaller value
of $\beta\sim1$ would appear to give a better fit.

The wavelength, $\lambda_0$, at which depolarization due to
stochastic Faraday rotation should occur follows
from equations (\ref{eq2.7}) and (\ref{eq2.8}):
\begin{eqnarray}
\label{eq4.3}
\lambda_0\approx30{\rm\,cm}
\times
\left(\frac{\kappa_0^{1/2}}
{10\rm\,rad\,m^{-2}} \right)
\left(\frac{\theta_B}
{10''} \right)^{-\beta/2},
\end{eqnarray}
where $\theta_B$ is the beamwidth of the telescope. With
$\kappa_0^{1/2}=10\rm\,rad\,m^{-2}$ for a
beamwidth of $\approx10''$, equation (\ref{eq4.3}) implies
depolarization at $\lambda\ga30{\rm\,cm}$. Thus, stochastic
Faraday rotation is plausible for the depolarization reported
by Lazio \et (1990). Such depolarization would be due to
variations in RM on a scale $r\la L\theta_B$, below the
resolution of the telescope.

\subsection{Depolarization in intergalactic media}

It has been suggested that differences in the degree of
polarization in the two lobes of radio galaxies might be due
to either internal depolarization or due to stochastic Faraday
rotation in a circumgalactic medium (Strom \& J\"agers
1988; Laing 1988; Garrington, Leahy \& Conway 1988). The observed
depolarization occurs typically at $\lambda$ between
$6\rm\,cm$ and $20\rm\,cm$. An example of an extragalactic
source for which detailed polarization data are available is
the inner $2\rm\,kpc$ of M87 (Owen, Eilek \& Keel 1990). This
source has relatively  high polarization, $\ga10$\%, in the
range $4.6$--$4.9\rm\,GHz$, with large values of
$\RM\sim1000$--$8000\rm\,rad\,m^{-2}$ that vary over scales
$\la1\rm\,kpc$ across the source.

Suppose that there are fluctuations in RM on a scale smaller
than can be observed, with variance LDM. Then the frequency
at which the depolarization should become significant is
$\nu_0\sim10{\rm\,GHz}\, [\LDM^{1/2}/1000\rm\,rad\,m^{-2}]$.
The absence of substantial depolarization indicates that
$\LDM^{1/2}$ cannot be much larger than
$1000\rm\,rad\,m^{-2}$. However, it would be surprising if
the magnetic field were so uniform that one had
$\LDM^{1/2}\ll\RM$, and hence one should expect
depolarization to become important at frequencies of order of
those already observed. Tribble (1991) used the theory for
beamwidth depolarization to place a limit on the value of $B$
in the circumgalactic medium.

The data for M87 should allow one to perform spatial
averages of $Q$, $U$ and $Q^2+U^2$ over maps at the
available frequencies, and this would provide a test for the
interpretation of the depolarization, which should affect
$Q$, $U$ but not $Q^2+U^2$. However, this test has yet to be
made.

\subsection{The Snake}

The Snake (Gray \et 1991) is a long ($\sim20'$), thin
($\la10''$) structure roughly perpendicular to the galactic
plane at $\sim1^\circ$ from the Galactic Centre. It is highly
polarized, $\sim60$\%, at $10.55\rm\,GHz$, and less
polarized at $4.79\rm\,GHz$ (Nicholls \& Gray 1992). There is
an abrupt decrease in polarization roughly at its mid point
(Gray 1993).

For the decrease in polarization to be  due to stochastic
Faraday rotation requires an abrupt increase in the
fluctuations in RM  at the point where the polarization
decreases. One possibility is that there is an interstellar
cloud along the line of sight covering only the weakly
polarized portion of the Snake. Testing this suggestion
requires multifrequency observations to determine RM along
the Snake.

\subsection{Pulsars}

Most pulsars are highly polarized (e.g., Lyne \& Manchester
1988), implying that stochastic Faraday rotation is normally
unimportant. However, there are exceptions which could provide
useful tests for the theory of stochastic Faraday rotation.

First consider the frequency at which depolarization might be
expected due to variations in RM in the ISM. The relevant
variations are over the apparent angular size, which along
with the time delay and the pulse width  may be derived
from a simple model for refractive scattering (Blandford \&
Narayan 1985, Rickett 1990). Gwinn, Bartel \& Cordes
(1993), found that the angular size is reasonably well fit by
an estimate based on the temporal broadening, $\tau$, through
$\theta_\tau=3.3(c\tau/L)^{1/2}$. At $327\rm\,MHz$ angular
sizes ranged from the detection limit($\sim2\rm\,mas$) to
$\sim65\rm\,mas$. Using equation (\ref{eq4.3}) with $\theta_B$ replaced
by the apparent angular size, $\la100\rm\,mas$, one concludes
that depolarization for most pulsars should occur only at
unobservably low frequencies, consistent with the
observations.

One exceptional case where depolarization of pulsars might be observable in a region around a supernova remnant (SNR) where
Alfv\'en waves are excited in association with shock acceleration of
relativistic particles (Spangler \et 1986).
Gwinn \et (1993) argued that such turbulence may account for
strong scattering around young SNR. These authors noted that
there are variations in DM with time up to
$0.01$--$0.06\rm\,pc\,cm^{-3}$. Assuming a magnetic field
$\sim3\rm\,\mu G$, this suggests the contribution to RM
from the turbulent region around the SNR could be comparable
to the total $\RM\sim100$--$1000\rm\,rad\,m^{-2}$. In a model
for SNR with sharp edges, Achterberg, Blandford
\& Reynolds (1994) suggested that the required Alfv\'en waves
imply fluctuations in the magnetic field
$\delta B/B\sim3\times10^{-3}$ on a scale $\sim10^{10}\rm\,m$,
which is smaller than the scattering disc. This model implies
fluctuations in RM across the scattering disc as large as
$\sim1\rm\,rad\,m^{-2}$, which suggests that depolarization
should occur at $\lambda\ga1\rm\,m$.

A clear example where depolarization does occur is for the
binary pulsar PSR B1259-63 which is eclipsed by the wind of
its companion for several weeks near periastron. Large
fluctuations in DM and RM were observed in the 1994 eclipse
(Johnston \et 1996), and, in the following eclipse in 1996,
depolarization was observed over a few days (around day
$-80$) before the pulsar disappeared and for several days as
it came out of eclipse (S.\ Johnston, private communication
1998). This depolarization was clearly associated with
fluctuations in RM on a time scale of minutes. This system provides the
possibility of testing the theory of stochastic
Faraday rotation in detail. Multifrequency observations of the
polarization on a short time scale in the epochs where the
variations in RM occur would provide a data set allowing the
averages of $Q$, $U$ and $Q^2+U^2$ to be performed over
various time scales as the level of fluctuations in RM
changes. It is hoped that such a data set will be recorded
around the next periastron.

\section{Conclusions}

It has long been known that stochastic Faraday rotation can
lead to depolarization (Burn 1966). Writing
$Q=Q_0\cos\phi_\F$, $U=Q_0\sin\phi_\F$, for the observed
Stokes parameters in terms of $Q_0$ at the source, and
assuming $\phi_\F=\phi_{\F0}+\delta\phi_\F$ with
$\delta\phi_\F$ obeying gaussian statistics, implies
$\langle Q\rangle=Q_0\cos\phi_{\F0}\,\exp[-\delta_l]$,
$\langle U\rangle=Q_0\sin\phi_{\F0}\,\exp[-\delta_l]$. It is
pointed out in the present paper that this depolarization
does not affect $\langle Q^2+U^2\rangle$, which should
remain constant while  $\langle Q\rangle^2+\langle U\rangle^2$
decreases. This effect, which is referred to as polarization
covariance, was first identified within the framework of a
more general theory for scintillations in an anisotropic
medium (paper~B), and stochastic Faraday rotation
provides an obvious interpretation of it. The interpretation
is that depolarization due to stochastic Faraday rotation
is due simply to random rotation of the Stokes vector. In
principle, whether depolarization is due to stochastic
Faraday rotation or to some other effect may be tested by
recording data on $Q$ and $U$ on the shortest time and
space scales available, and calculating $Q^2+U^2$ and
$\phi_\F$ from these data. In stochastic Faraday rotation,
$\phi_\F$ varies randomly while $Q^2+U^2$ remains constant.
The variance in the rotation measure is proportional to
$\langle\phi_\F^2\rangle$, but the same information is
provided by comparison of
$\langle Q\rangle^2+\langle U^2\rangle$ and
$\langle Q^2+U^2\rangle$. The latter quantities could be
recorded automatically on the shortest time scale available
on a telescope by modifying the software in polarization
measurements to record the value of $Q^2+U^2$, as well as the
values of $Q$ and $U$, so that the integrating over the time
during the observation gives $\langle Q^2+U^2\rangle$,
$\langle Q\rangle$ and $\langle U\rangle$ directly.

In principle, polarization covariance provides a definitive
test for depolarization due to stochastic Faraday rotation,
and the conditions under which this test should apply are
discussed in section~3. There are at least five relevant
averages:

\noindent
1.\ Beamwidth averaging: the average is over the beamwidth of
the observation. If the source is larger than the scattering
disc but smaller than the telescope beam, this average is
over the unresolved source.

\noindent
2.\ Averaging of a scattering-broadened image: the average is
over the scattering disc.

\noindent
3.\ Bandwidth averaging: the average is with respect to
frequency over the bandwidth of the observation.

\noindent
4.\ Spatial averaging: the average is with respect to
position over a polarization map.

\noindent
5.\ Temporal averaging: the average is over a temporal
sequence of observations.

Of these five, the first three are intrinsic to the
collection of the data, and polarization covariance does not
apply to them. However, for spatial and temporal averages,
stochastic Faraday rotation leads to a depolarization in
which $\langle Q\rangle^2+\langle U\rangle^2$ decreases and
$\langle Q^2+U^2\rangle$ remains constant.

Any observation of such depolarization provides information
on the statistical properties of fluctuations in the
magnetized plasma. Specifically, the wavelength at which the
depolarization sets in may be used to define a linear
depolarization measure, LDM, cf.\ equation (\ref{ldm}), which provides
a measure of the variance in RM, and hence on the fluctuations
in $(n_eB_\parallel)^2$ along the ray path. Available data on
the variation in RM in the ISM suggest that
$\langle B_\parallel^2\rangle/\langle B_\parallel\rangle^2$ is
slightly greater than unity (Simonetti \et 1984; Simonetti \&
Cordes 1988).

The definitive test for depolarization due to stochastic
Faraday rotation has yet to be applied to any specific data
set. Possible systems for which depolarization is plausibly
due to stochastic Faraday rotation are discussed in
section~5. Polarization maps of the lobes of radiogalaxies
could be used to test the theory for spatial averaging.
The next eclipse of the binary pulsar PSR B1259-63 should
provide an almost ideal test of the theory for temporal
averaging.


\section*{Appendix}

The theory of strong scintillations (e.g., Ishimaru 1978) may
be developed in terms of equations for the moments of the wave
amplitude. The wave amplitude, $u(z,{\bf r})$, is described
in the parabolic approximation; with the paraxial direction
along the
$z$ axis and with $\bf r$ the two-dimensional vector
perpendicular to this direction. The equation for the $n$th
order moment involves a first derivative with
respect to $z$ and second derivatives with respect the $n$
${\bf r}$s. The statistical average over the fluctuations is
based on an ensemble average assuming gaussian statistics. The
generalization to an anisotropic medium was made by Kukushkin
\& Ol'yak (1990, 1991) and in papers~A and~B). In a simple
plasma in which the modes are circularly polarized, the
dielectric tensor, $K_{ij}$, may be separated into an
anisotropic part, $K_\I$, and a part, $K_\V$, associated with
the anisotropy. Three correlation functions are needed to
describe the fluctuations, $\delta K_\I$, $\delta K_\V$:
these are written $A_{\X\Y}(z,{\bf r})=A_{\Y\X}(z,{\bf r})$,
which are proportional to
$\langle\delta K_\X(z,{\bf r})\delta K_\Y(z,{\bf0})\rangle$,
with $X,Y=I,V$. Alternatively, the fluctuations may be
described in terms of the phase structure function,
$D(z,{\bf r})=\langle[\phi(z,{\bf r})-\phi(z,{\bf0})]^2
\rangle$, with $D(z,{\bf r})=2\Delta z
[A(z,{\bf0})-A(z,{\bf r})]$ for a `thin screen' of thickness
$\Delta z$. The generalization to an anisotropic medium
involves three such functions, $D_{\X\Y}(z,{\bf r})$,
constructed from the three $A_{\X\Y}(z,{\bf r})$.
When considering the effects of Faraday rotation on the
correlation functions of $Q$ and $U$ one is interested in the case where
all the ${\bf r}$s are zero. In paper~A it was
shown that the nontrivial second order moments then give
$${\partial\over\partial z}\,
\pmatrix{\langle{\tilde Q}\rangle\cr
\langle{\tilde U}\rangle\cr}
+{\omega^2\over2c^2}\,A_{VV}
\pmatrix{1&0\cr
0&1\cr}
\pmatrix{\langle{\tilde Q}\rangle\cr
\langle{\tilde U}\rangle\cr}=0,
\eqno(A.1)$$
Integration of
$(A.1)$ over a slab of thickness $\Delta z$ reproduces
equation (\ref{eq2.8}) with
$$2\delta_l={\omega^2\over c^2}\,A_{VV}\,\Delta z.
\eqno(A.2)$$
The fourth order moments include the correlation functions
between the Stokes parameters. In paper~B it was shown that,
when all the ${\bf r}$s are zero, the nontrivial fourth order
moment equations then reduce to
{
\catcode`#\active
\setbox0=\hbox{$-$}\dimen0=\wd0
\def #{\hbox to \dimen0{\hfil}}
$$\displaylines{\hfill
{\partial\over\partial z}\,
\pmatrix{\langle{\tilde Q}{\tilde Q}\rangle\cr
\langle{\tilde U}{\tilde U}\rangle\cr
\langle{\tilde Q}{\tilde U}\rangle\cr}
+{\omega^2\over c^2}\,A_{VV}
\pmatrix{#1&-1&0\cr
\noalign{\vskip3pt}
-1&#1&0\cr
#0&#0&2\cr}
\pmatrix{\langle{\tilde Q}{\tilde Q}\rangle\cr
\langle{\tilde U}{\tilde U}\rangle\cr
\langle{\tilde Q}{\tilde U}\rangle\cr}=0,
\hfill
\llap{(A.3)}\cr}$$

\noindent where the tildes have the same meaning as in
section~2, and with $A_{VV}=A_{VV}(z,{\bf0})$. Integration of
$(A.1)$ over a slab of thickness $\Delta z$ reproduces equation
(\ref{eq2.11}) with $\delta_l$ given by $(A.2)$. This formally
justifies the treatment of the Faraday angle $\phi_\F$ as a
random variable in section~2 in discussing the correlation
functions between the Stokes parameters.


\section*{Acknowledgments}
We thank  Lawrence Cram, Ron Ekers, Simon Johnston, Jenny
Nicholls and Mark Walker for helpful comments.


\references

\reference
{Achterberg, A., Blandford, R.D., \& Reynolds, S.P.}
Achterberg, A., Blandford, R.D., \& Reynolds, S.P.,
{1994},
{A\&A, {281}, 220}

\reference
{Blandford, R., \& Narayan, R.}
Blandford, R., \& Narayan, R.,
{1985},
{MNRAS, {213}, 591}

\reference
{Burn, B.J.}
Burn, B.J.,
{1966},
{MNRAS, {113}, 67}

\reference
{Cordes, J.M., \& Lazio, T.J.}
Cordes, J.M., \& Lazio, T.J,
{1991},
{ApJ, {376}, 123}

\reference
{Cordes, J.M., Weisberg, J., Frail, D.A., Spangler, S.R.,
\& Ryan, M.}
Cordes, J.M., Weisberg, J., Frail, D.A., Spangler, S.R.,
\& Ryan, M,
{1991},
{Nature, {354}, 121}

\reference
{Dreher, J.W., Carilli, C.L., \& Perley, R.A.}
Dreher, J.W., Carilli, C.L., \& Perley, R.A.,
{1987},
{ApJ, {316}, 611}

\reference
{{Erukhimov, L.M., \& Kirsh, P.I.}
{1973}, {Izv.\ Vyssh.\ Uchebn.\ Zaved.\ Radiofiz}., {16}, 1783}
Erukhimov, L.M., \& Kirsh, P.I.,
{1973}, {Izv.\ Vyssh.\ Uchebn.\ Zaved.\ Radiofiz}., {16}, 1783

\reference
{Garrington, S.T., Leahy, J.P., Conway, R.G., \& Laing, R.A.}
Garrington, S.T., Leahy, J.P., Conway, R.G., \& Laing, R.A.,
{1988},
{Nature, {331}, 149}

\reference
{Gray, A.D.}
Gray, A.D.,
{1993},
{PhD thesis, University of Sydney}

\reference
{Gray, A.D., Cram, L.E., Ekers, R.D.,
\& Goss, W.M.}
Gray, A.D., Cram, L.E., Ekers, R.D.,
\& Goss, W.M.,
{1991},
{Nature, {353}, 237}

\reference
{Gwinn, C.R., Bartel, N., \& Cordes, J.M.}
Gwinn, C.R., Bartel, N., \& Cordes, J.M.,
{1993},
{ApJ, {410}, 673}

\reference
{Hamilton, P.A., Hall, P.J., \& Costa, M.E.}
Hamilton, P.A., Hall, P.J., \& Costa, M.E.,
{1985},
{MNRAS, {214}, 5P}

\reference
{Ishimaru, A.}
Ishimaru, A.,
{1978},
{Wave Propagation and Scattering in Random Media. Volume 2},
{Academic Press, New York}

\reference
{Johnston, S., Manchester, R.N., Lyne, A.G., D'Amico, N.,
Bailes, M., Gaensler, B.M., \& Nicastro, L.} Johnston, S., Manchester, R.N., Lyne, A.G., D'Amico, N.,
Bailes, M., Gaensler, B.M., \& Nicastro, L.,
{1996},
{MNRAS {279}, 1026}

\reference
{Kukushkin, A.V., \& Ol'yak, M.R.}
Kukushkin, A.V., \& Ol'yak, M.R.,
{1990},
{{Radiophys.\ Quantum Electron}.\ {\vf 33}, 1002}

\reference
{Kukushkin, A.V., \& Ol'yak, M.R.}
Kukushkin, A.V., \& Ol'yak, M.R.,
{1991},
{{Radiophys.\ Quantum Electron}.\ {\vf 34}, 612}

\reference
{Laing, R.A.}
Laing, R.A.,
{1988},
{Nature, {331}, 147}

\reference
{Lazio, T.J.,  Spangler, S.R. \& Cordes, J.M.}
Lazio, T.J.,  Spangler, S.R. \& Cordes, J.M.,
{1990},
{ApJ, {363}, 515}

\reference
{Lee, L.C., \& Jokipii, J.R.}
Lee, L.C., \& Jokipii, J.R.,
{1975},
{ApJ, {196}, 695}

\reference
{Lyne, A.G., \& Manchester, R.N.}
Lyne, A.G., \& Manchester, R.N.,
{1988},
{MNRAS, {234}, 477}

\reference
{Melrose, D.B. 1993a, J.\ Plasma Phys.,
{50}, 267}
Melrose, D.B. 1993a, J.\ Plasma Phys.,
{50}, 267

\reference
{Melrose, D.B. 1993b, J.\ Plasma Phys.,
{50}, 283}
Melrose, D.B. 1993b, J.\ Plasma Phys.,
{50}, 283

\reference{Narayan}
{Narayan, R., 1992, Phil. Trans. R. Soc. Lond. A, {341}, 151

\reference
{Nicholls, J., \& Gray, A.D.}
Nicholls, J., \& Gray, A.D.,
{1992},
{Proc.\ Aston.\ Soc.\ Australia, {10}, 233}

\reference
{Owen, F.N., Eilek, J.A., \& Keel, W.C.}
Owen, F.N., Eilek, J.A., \& Keel, W.C.,
{1990},
{ApJ, {362}, 449}

\reference
{Rohlfs, K., \& Wilson, T.L.}
Rohlfs, K., \& Wilson, T.L.,
{1996},
{Tools of radio astronomy}
{Springer, Berlin}

\reference
{Simonetti, J.H., \& Cordes, J.M.}
Simonetti, J.H., \& Cordes, J.M.,
{1986},
{ApJ, {310}, 160}

\reference
{Simonetti, J.H., \& Cordes, J.M.}
Simonetti, J.H., \& Cordes, J.M.,
{1988},
{in {Cordes, J.M., Rickett, B.J., \& Backer, D.C. (eds)},
{\it Radio Wave scattering in the Interstellar Medium}, p.\
134}

\reference
{Simonetti, J.H., Cordes, J.M.. \& Spangler, S.R.}
Simonetti, J.H., Cordes, J.M.. \& Spangler, S.R.,
{1984},
{ApJ, {284}, 126}

\reference
{Spangler, S.R.}
Spangler, S.R.,
{1982},
{ApJ, {261}, 310}

\reference
{Spangler, S.R., Mutel, R.L., Benson, J.M., \& Cordes,
J.M.}
Spangler, S.R., Mutel, R.L., Benson, J.M., \& Cordes,
J.M.,
{1986},
{ApJ, {301}, 312}

\reference
{Strom, R.G., \& J\"agers, W.J.}
Strom, R.G., \& J\"agers, W.J.,
{1988},
{A\&A, {194}, 79}

\reference
{{Tamoikin, V.V., \& Zamek, I.G.}
{1974}, {Izv.\ Vyssh.\ Uchebn.\ Zaved.\ Radiofiz}., {17}, 31}
Tamoikin, V.V., \& Zamek, I.G.,
{1974}, {Izv.\ Vyssh.\ Uchebn.\ Zaved.\ Radiofiz}., {17}, 31

\reference
{Taylor, J.H., \& Cordes, J.M.}
Taylor, J.H., \& Cordes, J.M.,
{1993},
{ApJ, {320}, L35}

\reference
{Tribble, P.C.}
Tribble, P.C.,
{1991},
{MNRAS, {250}, 726}

\endreferences

\end{document}